%% file: main.tex
\documentclass[journal]{IEEEtran}
\usepackage{caption}
\usepackage{algorithmic}
\usepackage[linesnumbered,ruled]{algorithm2e}
\usepackage{graphics,epstopdf}
\usepackage{balance}
\usepackage{listings}
\usepackage{comment}
\usepackage{enumitem}
\usepackage{xcolor}
\usepackage{fixmath}
\usepackage{cite}
\usepackage{amsmath,amssymb,amsfonts}
\usepackage{graphicx}
\usepackage{textcomp}
\usepackage{tabularx}
\usepackage{booktabs}
\usepackage{longtable}
\usepackage{multirow}
\usepackage{colortbl}
\usepackage{hhline}
\usepackage{tabularx,colortbl}
\usepackage{stmaryrd}
\usepackage{cite}
\usepackage{enumitem}
\usepackage{soul}
\usepackage{bm} 
\usepackage{mathtools} 
\usepackage{mathrsfs} 
\usepackage{multirow}
\usepackage{array}
\ifCLASSINFOpdf

\else

\fi

\begin{document}

\title{Efficient Privacy-Preserving Electricity Theft Detection with Dynamic Billing and\\Load Monitoring for AMI Networks}


\author{\IEEEauthorblockN{Mohamed I. Ibrahem,
Mahmoud Nabil,
Mostafa M. Fouda,~\IEEEmembership{Senior~Member,~IEEE,}\\
Mohamed Mahmoud,~\IEEEmembership{Member,~IEEE,}
Waleed Alasmary,~\IEEEmembership{Senior~Member,~IEEE,} and \\ Fawaz Alsolami,~\IEEEmembership{Member,~IEEE}}
\thanks{Corresponding author: Mohamed I. Ibrahem.}
\thanks{M. I. Ibrahem, and M. Mahmoud are with the Department of Electrical and Computer Engineering, Tennessee Tech. University, Cookeville, TN 38505 USA (e-mail: miibrahem42@students.tntech.edu; mmahmoud@tntech.edu).}
\thanks{M. Nabil is with the Department of Electrical and Computer Engineering, North Carolina A and T University, USA (e-mail: mnmahmoud@ncat.edu).}
\thanks{M. M. Fouda is with the Department of Electrical and Computer Engineering, Tennessee Tech. University, Cookeville, TN 38505 USA, and the Department of Electrical Engineering, Faculty of Engineering at Shoubra, Benha University, Egypt (e-mail: mfouda@ieee.org).}
\thanks{W. Alasmary is with the Department of Computer Engineering, Umm Al-Qura University, Saudi Arabia (e-mail: wsasmary@uqu.edu.sa).}
\thanks{F. Alsolami is with the Department of Computer Science, King Abdulaziz University, Saudi Arabia (e-mail: falsolami1@kau.edu.sa).} \vspace{-0.3cm}
}

\maketitle

\markboth{Ibrahem \MakeLowercase{\textit{et al.}}: Privacy-Preserving Electricity Theft Detection with Dynamic Billing and Load Monitoring for AMI Networks}%
{}
\IEEEpeerreviewmaketitle
\input{Files/abstract.tex}


\input{Files/Introduction.tex}
\input{Files/RelatedWork.tex}

\input{Files/SystemModels.tex}
\input{Files/Preliminaries.tex}
\input{Files/ProposedScheme.tex}
\input{Files/Security.tex}
\input{Files/Conclusion.tex}

\newcolumntype{P}[1]{>{\centering\arraybackslash}p{#1}}
\bibliographystyle{IEEEtran}
\bibliography{main}

\end{document}

%% file: Files/abstract.tex
\begin{abstract}
In advanced metering infrastructure (AMI), smart meters (SMs) are installed at the consumer side to send fine-grained power consumption readings periodically to the system operator (SO) for load monitoring, energy management, billing, etc. However, fraudulent consumers launch electricity theft cyber-attacks by reporting false readings to reduce their bills illegally. These attacks do not only cause financial losses but may also degrade the grid performance because the readings are used for grid management. To identify these attackers, the existing schemes employ machine-learning models using the consumers’ fine-grained readings, which violates the consumers’ privacy by revealing their lifestyle. In this paper, we propose an efficient scheme that enables the SO to detect electricity theft, compute bills, and monitor load while preserving the consumers’ privacy. The idea is that SMs encrypt their readings using functional encryption, and the SO uses the ciphertexts to ($i$) compute the bills following dynamic pricing approach, ($ii$) monitor the grid load, and ($iii$) evaluate a machine-learning model to detect fraudulent consumers, without being able to learn the individual readings to preserve consumers’ privacy. We adapted a functional encryption scheme so that the encrypted readings are aggregated for billing and load monitoring and only the aggregated value is revealed to the SO. Also, we exploited the inner-product operations on encrypted readings to evaluate a machine-learning model to detect fraudulent consumers. Real dataset is used to evaluate our scheme, and our evaluations indicate that our scheme is secure and can detect fraudulent consumers accurately with low communication and computation overhead.

\end{abstract}


%% file: Files/Introduction.tex
{\section{Introduction}\label{sec:introduction}}


\begin{figure*}[t]
\centering
\includegraphics[width=0.8\textwidth]{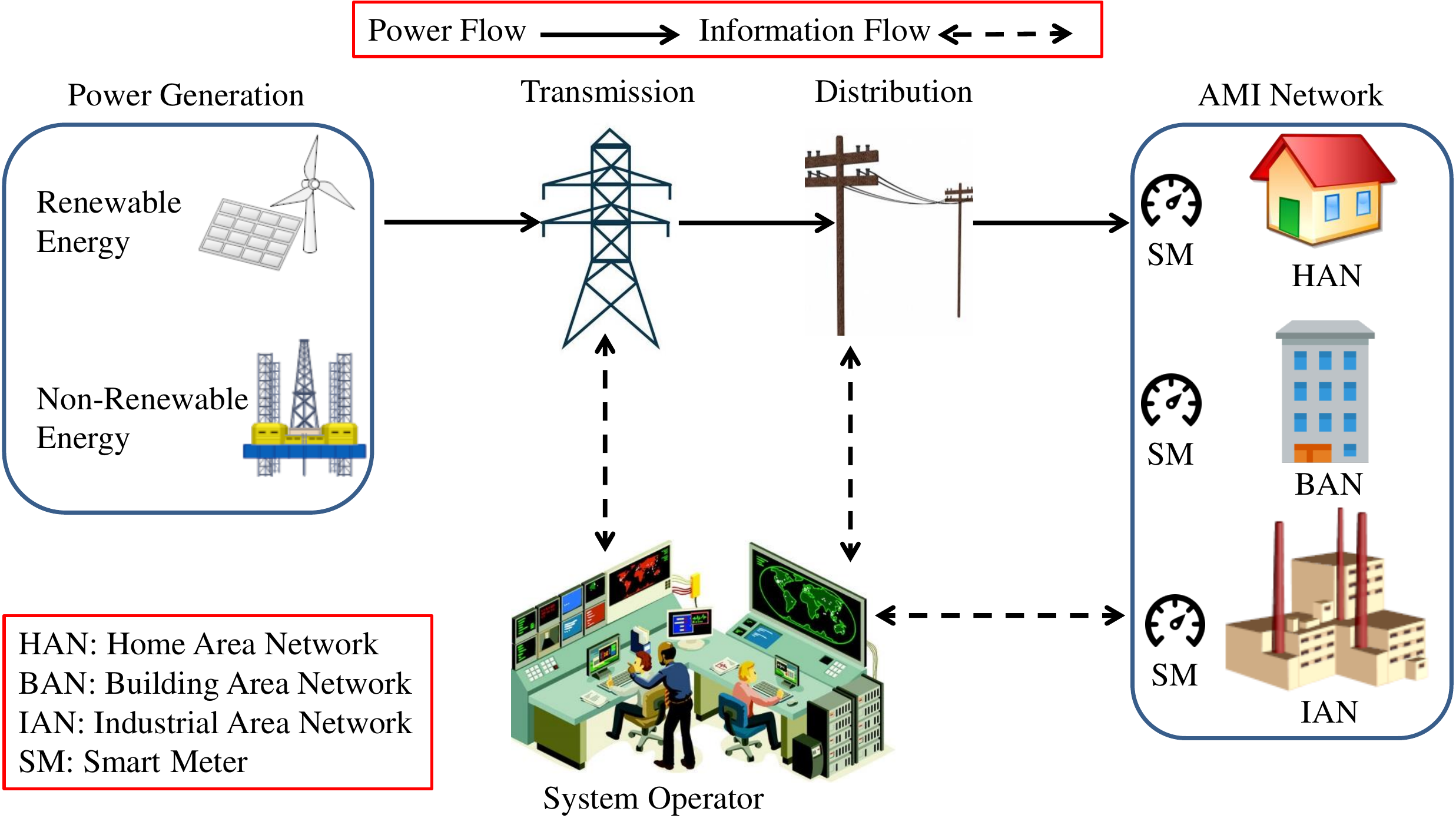}
\caption{Smart grid conceptual architecture. \label{fig:SG_arch}}
\end{figure*}

\IEEEPARstart{S}{mart} grid (SG) is an advanced upgrade to the traditional power grid that aims to facilitate reliable delivery of electricity, optimize grid operation, and engage consumers~\cite{6298960}. 
Fig.~\ref{fig:SG_arch} illustrates the model structure of the SG, which comprises of advanced metering infrastructure (AMI) network, electricity generation sources, transmission and distribution systems, and a system operator (SO).
AMI enables the bi-directional communication between the smart meters (SMs), which are deployed at consumer premises, and SO for regular load monitoring, energy managment, and billing~\cite{5741147}.
Unlike the traditional power grid that collects the power consumption readings monthly, AMI network collects fine-grained power consumption readings (every few minutes) measured/sent by SMs. Then, these readings are forwarded to the SO for monitoring the load, controlling the energy supply efficiently, and calculating the consumers' bills. These bills follow dynamic pricing approach in which the tarrif of electricity consumption changes through the day to stimulate consumers to reduce consumption during peak hours~\cite{alsharif2018epic}.


In SG, electricity theft attacks can be launched by fraudulent consumers who tamper with their SMs so that they  report lower consumption readings to reduce energy bill illegally. This deceptive behaviour does not only cause financial losses, but also the false readings used for load monitoring may affect the decisions made by the SO regarding grid management, which may cause the instability of the grid or blackout in severe cases~\cite{8373734}. 
Electricity theft is a serious problem in the existing power grid that causes hefty financial losses. For instance, the United States loses about \$6 billion annually due to electricity thefts~\cite{ETSIB}. 
The losses in developing countries also have extremely bad consequences. For example, India suffers from about $\$17$ billion losses every year because of electricity theft~\cite{jokar2016electricity}.

In order to identify the fraudulent consumers, machine learning-based models, which are trained on fine-grained power consumption readings, have been proposed~\cite{zheng2018wide,issue4,jokar2016electricity}. 
However, revealing the consumers' fine-grained power consumption readings to the SO for electricity theft detection, load monitoring, and billing creates a serious privacy problem. This is because the fine-grained readings expose the consumers' life habits, whether they are at home or on-leave, number of people at home, the appliances they are using, etc~\cite{issue1}. This may result in criminal activities, e.g., thieves can break into homes when consumers are absent~\cite{issue4}. On the other side, these private data may be sold to insurance companies to adapt their plans based on the consumers' activities.
In summary, the research problem we address in this paper is \textit{how to enable the SO to monitor load, compute bills, and detect fraudulent consumers without learning the fine-grained power consumption readings} of the consumers to preserve their privacy.





In the literature, the proposed scheme in \cite{8746794} ``PPETD'' tried to address this research problem.
It uses secret sharing technique to allow sending the fine-grained power consumption readings in a masked manner in such a way that the SO can obtain the aggregated readings for billing and load monitoring without being able to learn the individual readings to preserve consumer privacy.
It also employs a convolutional neural network (CNN) machine learning model based on secure multi-party computation protocols using arithmetic and binary circuits. These protocols are executed interactively by the SO and each SM to evaluate the CNN model on the reported masked fine-grained power consumption readings without learning the readings to preserve the consumers' privacy. 
However, this scheme suffers from the following issues.
\begin{enumerate}
    \item The computation and communication overheads are impractically high. The total time needed for the CNN model evaluation using masked readings is around 48 minutes and the amount of exchanged data is 1900 MB. This may be impractical for SMs because cost-effective devices tend to have limited computation capability and low bandwidth communication. Moreover, this evaluation is done in an online and interactive communication session between each SM and the SO. Obviously, the requirement of a long session between each SM and the SO and exchanging much amount of data is not practical, scalable, or even cost-effective given that cellular communication may be used to enable the communications between the SMs and SO.
    \item There is a trade-off between the accuracy of the model and overhead due to the approximation done for a non-linear function (sigmoid function).
    \item The classification of the model is known to both SM and SO, which is supposed to be known only to the SO. By knowing the classification of the model, the fraudulent consumer can return the original software to the SM before the SO sends technicians to inspect it to avoid liability.
\end{enumerate}

Therefore, in this paper, we address these limitations by proposing a privacy-preserving and efficient
\textbf{e}lectricity \textbf{t}heft \textbf{d}etection scheme enabling dynamic billing and load monitoring using \textbf{f}unctional \textbf{e}ncryption (FE), named ``\textbf{ETDFE}''. 
The idea is that the SMs encrypt their fine-grained readings using functional encryption scheme and send the ciphertexts to the SO.
We adapted the functional encryption scheme~\cite{10.1007/978-3-319-96884-1_20} to enable aggregating the SMs' encrypted readings, and revealing only the aggregated readings to the SO for billing and load monitoring without being able to learn the individual readings to preserve consumers' privacy.
Furthermore, we train a deep learning-based electricity theft detection model and leverage the inner product operations on encrypted data supported by the FE to evaluate the model using the encrypted fine-grained readings without revealing the readings to the SO to preserve privacy. 

Using real dataset, we evaluated the performance of our electricity theft detection model. We also analyzed the security of our scheme and measured the communication and computation overhead. Our evaluations confirm that our scheme is secure and can detect electricity thefts accurately with much less communication and computation overhead comparing to the scheme proposed in~\cite{8746794}.
Specifically, our scheme can significantly reduce the computation and communication overheads.
Moreover, unlike~\cite{8746794}, our proposed scheme does not need both SMs and the SO to involve in online/interactive session to evaluate the electricity theft detection model.

The remainder of this paper is organized as follows. Section~\ref{sec:Related Work} discusses the related works. Then, our system models and design objectives are discussed in section~\ref{sec:System Models}. Section~\ref{sec:Preliminaries} illustrates the preliminaries used in our work. Our envisioned ETDFE scheme is presented in section~\ref{sec:Proposed Scheme}. Next, The performance evaluation and security analysis of our scheme are discussed in sections~\ref{sec:Performance Evaluations} and~\ref{sec:Security Analysis}, respectively. Finally, the paper is concluded in section~\ref{sec:Conclusions}.

%% file: Files/RelatedWork.tex
\section{Related Work} \label{sec:Related Work}

A few works in the literature have tried to address privacy-preserving electricity theft detection in SG \cite{8746794,salinas2012privacy,yao2019energy,salinas2013privacy,salinas2016privacy}. Some schemes are based on machine learning techniques~\cite{8746794,yao2019energy}, while others use different techniques~\cite{salinas2012privacy,salinas2013privacy,salinas2016privacy}.

The work done by Salinas et. al. \cite{salinas2012privacy,salinas2013privacy} have attempted to investigate the privacy issue in detecting electricity theft.
They proposed three distributed peer-to-peer (P2P) computing algorithms based on lower–upper decomposition (LUD) to preserve privacy. Such algorithms solve linear system of equations (LSE) for the consumers' ``honesty coefficients'' to detect fraudulent consumers who commit energy theft. After a mutual communication between the SMs to solve LSE, the SO receives the honesty coefficient from each SM. If the honesty coefficient is equal to one, this consumer is honest, otherwise, the consumer reported less power consumption.
Although this scheme can successfully identify all the energy thieves in a small size network, it may be unstable in large networks due to the rounding errors in LU decomposition.
In addition, the scheme fails if the SMs tamper with the messages sent to other parties. Furthermore, the power line losses are assumed to be known, which are difficult to acquire practically. 
Besides, this scheme takes into consideration only one type of attack in which the fraudulent consumers reduce their power consumption reading with constant reduction rates, where the real consumption readings are multiplied by a constant number that is less than one. However, there are many other energy theft scenarios, such as by-pass filters~\cite{jokar2016electricity}. Finally, the scheme does not consider load monitoring and dynamic billing.


The electricity theft detection scheme presented in \cite{salinas2016privacy} considers consumers' privacy by using Kalman filter-based P2P state estimation protocol to find the line currents and biases of the consumers. 
The main idea of this scheme is to use state estimation techniques by the SO to identify the fraudulent consumers after receiving estimations of line segment currents and biases from all SMs. The SMs with biases larger than a predefined threshold are considered fraudulent.
The privacy of this scheme is guaranteed by employing a distributed kalman filter, where the SO does not need to access the consumers' power consumption readings.
However, this work significantly varies from ours in three perspectives. 
First, we use a machine learning model to determine electricity thefts, which usually performs better than state estimation approaches \cite{jokar2016electricity}. 
Second, the proposed state estimator is based on a set of distributed algorithms executed by SMs, and hence, the scheme may fail if SMs tamper with the messages sent to other peers.
Last, our scheme enables dynamic billing and load monitoring, which are not considered in \cite{salinas2016privacy}.

Machine learning-based models have been proposed in~\cite{8746794,yao2019energy} to identify electricity thefts. A CNN model is used in \cite{yao2019energy} to detect fraudulent consumers. In this scheme, SMs send their encrypted electricity consumption readings to two system entities. One entity, which is assumed to be fully trusted, is responsible for running a CNN model (i.e., electricity theft detector) after decrypting the consumer's fine-grained readings, and then reports the output of the model to the SO. 
Another entity, which is assumed distrusted, aggregates the consumers' encrypted power consumption readings in a certain residential area to obtain the aggregated reading for load monitoring without being able to learn the individual readings to preserve privacy.
Practically, it is difficult to ensure that an entity would not abuse consumers' information; in addition, this scheme cannot support dynamic billing.


\begin{figure*}[t]
\centering
\includegraphics[width=0.8\textwidth]{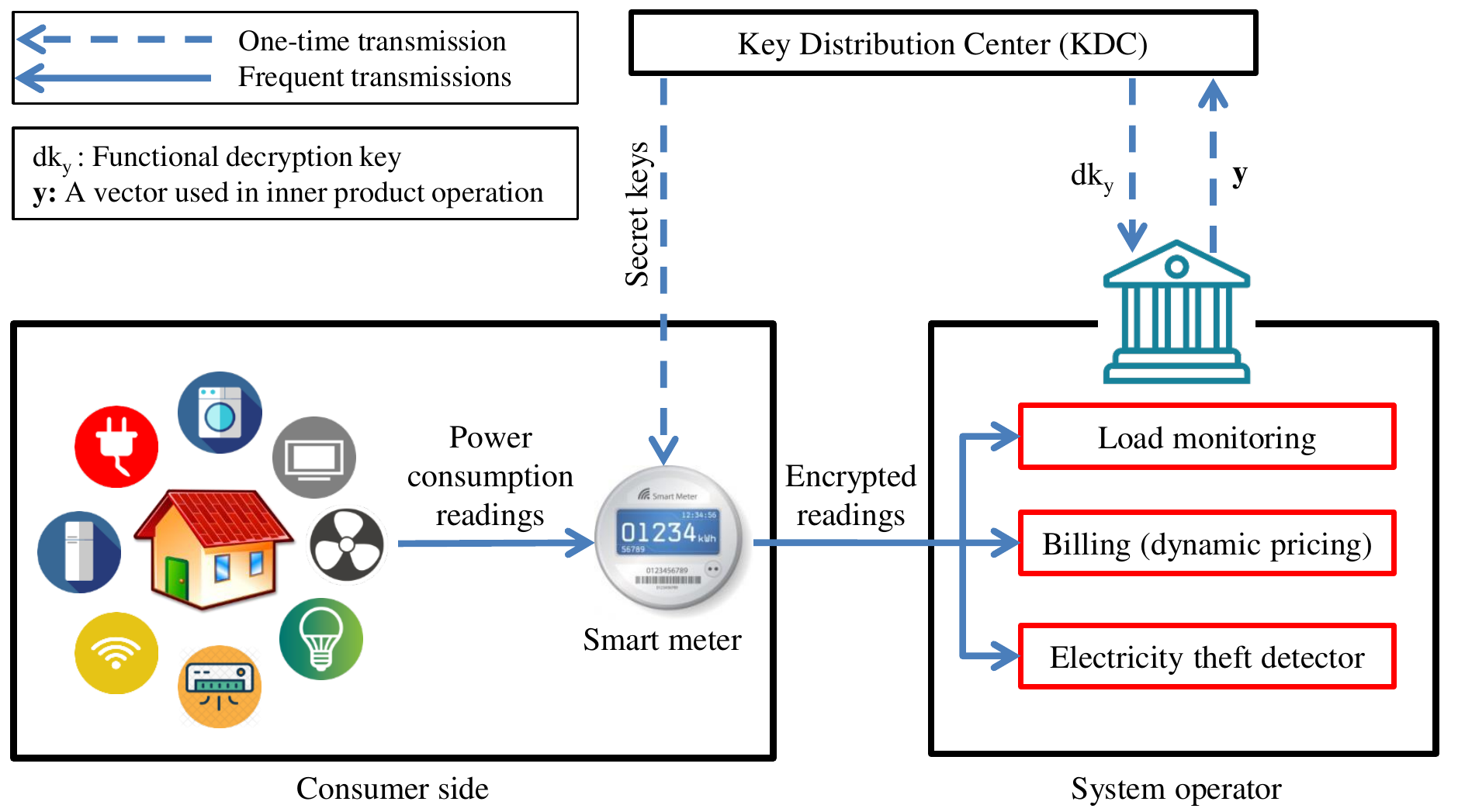}
\caption{Network Model.} \label{fig:network_model}
\end{figure*}


Nabil et. al.~\cite{8746794} have proposed a privacy-preserving scheme that enables the SO to detect fraudulent consumers, who steal electricity, by developing a CNN machine learning model based on secure multi-party computation protocols using arithmetic and binary circuits. These protocols are executed by the SO and each SM in an online/interactive session to evaluate the CNN model using the reported masked fine-grained power consumption readings. The proposed scheme uses also secret sharing technique to share secrets allowing SMs to send masked readings to the SO such that these readings can be aggregated for the purpose of monitoring and billing. The scheme also enables billing using dynamic pricing rates in which the tariff of the electricity changes during the day to stimulate consumers to not use electricity in peak hours to reduce the demand. However, the scheme suffers from the following drawbacks.
\begin{enumerate}
    \item The proposed scheme requires high computation and communication overhead. The SMs and SO should run a machine learning model in an interactive way (i.e., online) to maintain the consumers' privacy while allowing the SO to detect whether a consumer is honest or fraudulent. 
    Furthermore, to evaluate the model for a single SM, the total time needed is around 48 minutes and the amount of exchanged data is 1900 MB. The scheme also requires another overhead for running a technique to share the secrets needed to mask the readings. 
    This large computation and communication overheads are impractical for SMs because cost-effective devices tend to have limited computation capability and low bandwidth communications.
    \item A non-linear function (sigmoid) is used in the model and in order to evaluate the function on masked readings, it is approximated as a linear function. This creates a trade-off between the accuracy of the model and overhead, i.e., the better the approximation, the better the accuracy but with more overhead.
    \item The classification of the model is known to both SM and SO, which is supposed to be known only to the SO. By knowing the classification of the model, the fraudulent consumer can return the original software to the SM before the SO sends technicians to inspect it to avoid liability.
\end{enumerate}

%% file: Files/SystemModels.tex
\section{System Models and Design Objectives}  \label{sec:System Models}

This section discusses the considered network and threat models as well as the design objectives of our scheme. 

\subsection{Network Model} \label{subsec:Network Model} 
As shown in Fig.~\ref{fig:network_model}, our considered network model includes the consumer-side (smart meters), system operator-side, and an offline key distribution center (KDC). The role of each entity is described below.
\begin{itemize}
    \item \textit{Smart Meter (SM)}: 
    The consumer has smart appliances at his/her home which are connected to the SM. Each SM sends its fine-grained power consumption readings periodically (e.g., every 30 minutes) to the SO. A set of SMs, $\mathbb{SM}=\{SM_i, 1 \leq i \leq |\mathbb{SM}|\}$, form an AMI network. The SMs can communicate directly with the SO or they can communicate with the SO via a gateway. In the latter case, the SMs may communicate directly with the gateway, or multi-hop data transmission is used to connect the SMs to the gateway, where some SMs may act as routers to relay other SMs' data.
    
    \item \textit{System Operator (SO)}:
    The SO uses the fine-grained power consumption readings sent by SMs for load monitoring and energy management. Moreover, the SO uses these readings to evaluate a neural network model to detect electricity thefts and compute the bill for each consumer following dynamic pricing approach in which the electricity price increases at peak hours to stimulate consumers to reduce demand in these hours.
    
    \item \textit{Key Distribution Center (KDC)}: It distributes the public parameters in addition to the private keys, i.e., the encryption and functional decryption keys for both SMs and SO, respectively. KDC can be operated by a national authority such as the Department of Energy.

\end{itemize}


\subsection{Threat Model}\label{subsec:Threat Model} 
The SO may attempt to use the consumers' fine-grained power consumption readings to learn sensitive information including the consumers' activities, e.g., learning whether a consumer is at home or on-leave, and so forth. For consumers, they may conduct the following misbehaviour. First, they may send to the SO false (low) power consumption readings to reduce their bills illegally, which does not only cause financial losses but it may also result in wrong decisions regarding energy management. Second, the consumers may be interested in learning the fine-grained power consumption of other consumers to infer sensitive information about the lifestyle of the consumers. Regarding collusion, the SO may collude with consumer(s) to infer the readings of other consumers, but the number of colluding consumers should be fewer than or equal ($n-1$), where $n$ is the number of SMs. Moreover, some consumers may collude with others to infer sensitive information.

Basically, the objective of this paper is to preserve the consumers' privacy while using their fine-grained power consumption readings for load management, billing, and theft detection, i.e., no one including the SO should be able to learn the fine-grained readings of individual consumers.


\subsection{Design Objectives} \label{sub:design_requirements}	
Our scheme should achieve the following functionality and security requirements:

\subsubsection{Functionality Requirements}  \label{subsec:Functional Requirements} 

\begin{itemize}[leftmargin=1.25cm]
	\item [(F1)] In an AMI network, ETDFE should enable the SO to obtain the total electricity consumption of the consumers at each reporting period for load monitoring and energy management.

    \item [(F2)] Regarding billing, ETDFE should allow the SO to compute each consumer's electricity bill efficiently following dynamic pricing. 
    
   	\item [(F3)] ETDFE should allow the SO to run an electricity theft detector for each consumer using his/her fine-grained power consumption readings to detect whether this consumer is fraudulent or not.
 
\end{itemize}

\subsubsection{Security and Privacy Requirements} \label{subsec:Security Requirements} 

\begin{itemize}[leftmargin=1.25cm]
    \item [(S1)] Our electricity theft detector should be secure against any misbehaviour from fraudulent consumers who aim at stealing energy without being detected.
    
    \item [(S2)] Preserving consumers' privacy: No entity (including the SO) should learn the fine-grained power consumption readings of individual consumers at any reporting period.
    
    \item [(S3)] Confidentiality of AMI's total power consumption and  consumers' bills: SO should be the only entity that learns the total power consumption of all consumers in an AMI for load monitoring and the billing amount of each consumer as well.
    
\end{itemize}


%% file: Files/Preliminaries.tex


\section{Preliminaries} \label{sec:Preliminaries} 
\subsection{Functional Encryption}
\textit{Functional encryption} (FE) is a new cryptosystem that allows the encryptor to encrypt a message $\boldsymbol{x}$ using an encryption key, and enables the decryptor to perform computations on the encrypted message to learn the output of a predefined function $f$($\boldsymbol{x}$) using a functional decryption key without being able to learn the message $\boldsymbol{x}$ itself~\cite{10.1007/978-3-642-19571-6_16}.
Recently, the focus on FE has been increasing, especially how to design efficient schemes for limited classes of functions or polynomials, such as linear~\cite{10.1007/978-3-662-53015-3_12,10.1007/978-3-662-46447-2_33} or quadratic~\cite{10.1007/978-3-319-63688-7_3}. In this paper, we focus on the inner product functional encryption (IPFE) that allows to perform inner product operation over encrypted vector. 
In an IPFE scheme, given the encryption of a vector $\boldsymbol{x}$, and a functional decryption key associated with a vector $\boldsymbol{y}$, one can obtain only the dot product result $(\boldsymbol{x} \cdot \boldsymbol{y})$ by decrypting the encryption of $\boldsymbol{x}$ and without being able to learn $\boldsymbol{x}$. IPFE consists of three parties as follows.
\begin{itemize}
    \item \textit{Key Distribution Center (KDC)}: This generates the encryption and functional decryption keys for both the encryptor and decryptor, respectively.
    
    \item \textit{Encryptor}: It encrypts the plaintext vector $\boldsymbol{x}$ using the encryption key and sends the ciphertext to the decryptor.

    \item \textit{Decryptor}: It receives a functional decryption key $dk_y$ from the KDC, which is associated with a vector $\boldsymbol{y}$, and evaluates the dot product on the encrypted vector received from the encryptor. It has access only to the result of that dot product evaluation $(\boldsymbol{x} \cdot \boldsymbol{y})$, and of course, it must not collude with KDC. 
    
    \end{itemize}






\subsection {Feed-Forward Neural Networks (FFNs)}
FFNs are widely used in solving many challenging machine learning problems such as system identification of a biochemical process~\cite{170564}, face recognition system~\cite{6549322}, and age identification from voice~\cite{8404322}. This wide adoption of FNNs is due to their high accuracy.
FFNs are called feed-forward because the information only travels forward in the neural network, from the input nodes and through the hidden layer(s) (single or many layers) and finally through the output nodes. They are also called deep networks, multi-layer perceptron (MLP), or simply neural networks~\cite{10.5555/1213811}. 

\begin{figure}[t]
\centering
\includegraphics[width=3.4in]{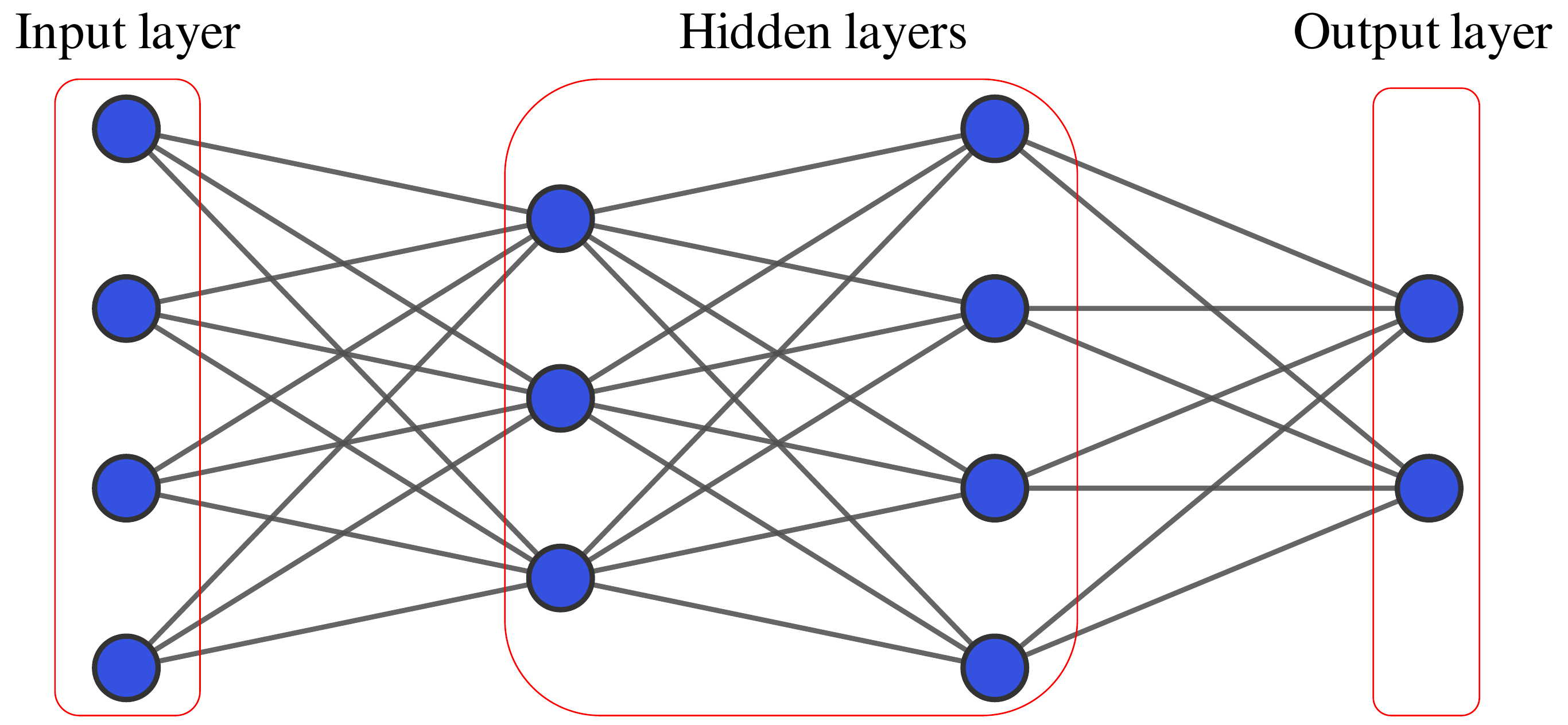}
\caption{Typical architecture of a feed forward neural network (FFN).} \label{fig:ffn_arch}
\end{figure}

Fig.~\ref{fig:ffn_arch} shows a typical architecture of an FFN that consists of:
\begin{itemize}
    \item \textit{Input Layer}: This is the first layer of a neural network. It consists of nodes, called neurons, that receive input data and pass them to the following layers. The number of neurons in the input layer is equal to the number of attributes or features of the input data.

    \item \textit{Output Layer}: This is the last layer which gives the prediction (or classification) of the model. The activation function used in this layer depends on the problem. For example, in a binary classifier, the output is either 0 or 1, and thus, a sigmoid activation function is usually used, while for a multi-class classifier, a softmax function is commonly used. On the other hand, for a regression problem, where the output is not a predefined category, we can simply use a linear activation function.
    
    \item \textit{Hidden Layers}: Between the input and output layers, there are hidden layer(s) that depend on the type of the model, e.g., the hidden layers of a CNN model typically consist of convolutional layers, pooling layers, etc. Hidden layers contain a vast number of neurons which apply transformations to the inputs before passing them. Every neuron in a layer is connected to all the neurons in the previous layer, and each connection may have a different strength or weight. When the network is trained, the weights are computed and updated in the direction of improving the model accuracy. By having multiple hidden layers, we can compute complex functions by cascading simpler functions. The number of hidden layers is termed as the depth of the neural network.
    
\end{itemize}


For a given neuron, the inputs are multiplied by the weights and summed together. This value is referred to as the summed activation of the neuron. The summed activation is then transformed via an activation function and defines the specific output or “activation” of that neuron.

In this paper, we use the FFN to solve a binary classification problem, i.e., to detect whether the consumer is honest or fraudulent. In machine learning, classification is a type of supervised learning method, where the task is to divide the data samples into predefined groups by a decision function. In the following, we discuss the training process of an FFN and the widely used activation functions. 

\subsubsection{FFN Training}
The features/input data is fed into the first layer of a neural network (i.e., input layer). Then, these features are gradually mapped to higher-level abstractions via the iterative update (a.k.a, feed-forward and back-propagation) in
the hidden layers of the neural network for a predefined number of iterations. These mapping abstractions, known as learned neural network model, can be used to predict the label in the output layer.

The training of such a network is quite complicated, when there exists an output error because it is hard to know how much error comes from the neurons and how to adjust the weights and biases~\cite{6745416}. Thus, the FNN training involves adjusting the weight and the bias parameters $\Theta$ by defining a cost function and selecting an optimizer. 
The problem can only be solved by finding the effect of all the weights in the network. This is done by the back-propagation algorithm~\cite{6745416} in which the FNN weights are updated using the gradients of the cost function with respect to the neural network's weights. In an FFN, the output values are compared with the correct prediction for optimizing the cost function. Then, the error is fed back through the network to adjust the weights of each connection in order to reduce the cost (loss) function~\cite{6745416}. For the cost function, \textit{categorical cross-entropy} $C(y,\hat{y})$ is defined to measure the loss due to the difference of two distributions, true distribution $y$ and learned distribution $\hat{y}$, for $M$ classes as follows: 


\begin{equation*}
    \begin{aligned}
        C(y,\hat{y}) = \underset{\Theta}{\min} (-\sum _{c=1}^{M} y(c)\ log(\hat{y}(c))).\label{eq:cross-entropy}
    \end{aligned}
\end{equation*}

During training, an optimization method, e.g., \textit{ADAM}~\cite{kingma2019method}, is used for optimizing the cost function. Supervised labeled data are used to train the neural network. In addition, hyper-parameters of the neural network such as the number of neurons in each layer, the number of layers, type of the optimizer, etc., can be determined using hyperopt python library~\cite{Bergstra_2015}, k-fold cross validation, or any other validation method \cite{8545748}.



\subsubsection{Activation Functions}
In a neural network, the activation function is responsible for transforming the summed weighted input from the neuron into the activation of that neuron. In the following, we explain some common activation functions and their usage.

\begin{itemize}
    \item \textit{Rectified Linear Unit} (ReLU): It allows positive values to pass through it, and maps negative values to zero. The main advantage of ReLU is the computational simplicity because it only requires a simple max() function as follows~\cite{nwankpa1811activation}. 
    \begin{align*}
        ReLU(x) & = max(0,x).
    \end{align*} 
    Unlike the tanh and sigmoid activation functions that use exponential operations, ReLU mostly acts like a linear activation function, and it is usually easier to optimize the neural network when its behavior is linear or close to linear.

    
    \item \textit{Softmax}: It is often used in the output layer for multi-class classification problems.  Softmax outputs a probability vector for a given input vector, i.e., for an input vector $\mathbf {z}=[\mathbf {z}[1],\ldots ,\mathbf {z}[M]]\in \mathbb {R} ^{M}$ of length $M$, where $M$ is the number of classes, the softmax function is defined as follows~\cite{nwankpa1811activation}.

\begin{equation*}
    Softmax ({z}[i])={\frac {e^{z[i]}}{\sum _{j=1}^{M}e^{z[j]}}} \  \ for \ \ i = \{1, \dots, M\}.
     \label{eq:1}
\end{equation*}
\end{itemize}


%% file: Files/ProposedScheme.tex

\section{Proposed Scheme} \label{sec:Proposed Scheme}
In this section, we first give an overview for the proposed ETDFE and then discuss system initialization, how SMs report their power consumption readings, and how the SO computes the aggregated readings for load monitoring. Next, we explain how the electricity bills are computed following dynamic pricing approach. Finally, we explain the way we train a machine learning model for electricity theft detection and discuss how the SO can use the SMs' encrypted readings to evaluate the model to detect electricity theft without learning the readings to preserve the consumers' privacy.

\subsection{Overview}
\begin{figure}[t]
\centering
\includegraphics[width=3.4in]{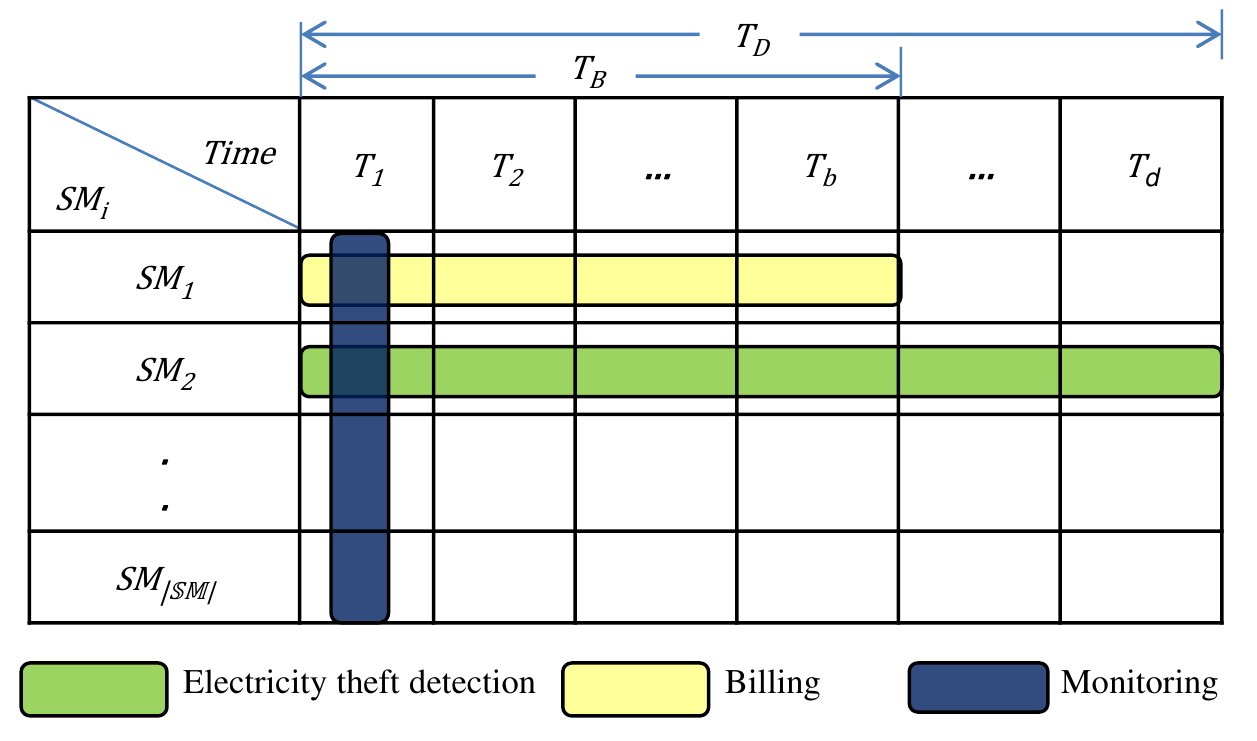}
\caption{Monitoring, billing, and electricity theft detection intervals.} \label{fig:System_scenario}
\end{figure}

The main phases of our scheme can be summarized as follows.
\begin{itemize}
    \item Using an FE scheme, each $SM_i \in \mathbb{SM}$ sends its encrypted readings periodically to the SO using the secret key $\boldsymbol s_{i}$ every time slot $T_t$ as shown in Fig.~\ref{fig:System_scenario}. 
    \item At every time slot $T_t$, the SO receives all encrypted readings from all SMs and uses the monitoring functional decryption key $\boldsymbol{dk_{m}}$ to obtain the aggregated reading of SMs in an AMI network for load monitoring without being able to learn the individual readings to preserve consumers' privacy.
    \item Regarding billing, as shown in Fig.~\ref{fig:System_scenario}, after receiving $b$ encrypted readings from each SM (which represent the readings per billing period $\mathit{T_B}$, where $\mathit{T_B}$= $\{\mathit{T_1},\mathit{T_2},\dots,\mathit{T_b}\}$), the SO applies dynamic pricing on these readings to compute the bill for each consumer $\boldsymbol{i}$ using the billing functional decryption key $DK_{bi}$ without learning the individual readings to preserve privacy.
    \item After receiving $d$ encrypted readings from the SMs, the SO uses the functional decryption key $DK_{di}$ of each $SM_i$ to evaluate an electricity theft machine learning model to detect whether this consumer is honest or fraudulent without learning the readings to preserve privacy.
\end{itemize}
For better readability, we define the main notations used in this section in Table~\ref{tab:multi_notation}. 

\subsection{System Initialization}

In system initialization\footnote{We use the standard lowercase notation for elements in $\mathbb{Z}_{q}$ and uppercase notation for elements in $G$.}, the KDC\footnote{\textit{KDC is needed only to bootstrap the system by distributing the necessary keys. After that, the system is run without involving it}}
should compute and distribute the following:
(1) Public parameters;
(2) SMs' encryption keys;
and (3) SO's functional decryption keys.


\subsubsection{Public Parameters}
To generate the public parameters, the KDC should: 
        \begin{itemize}
            \item generate $\{\, \mathbb{G}, q, P\}\,$ where $\mathbb{G}$ is a cyclic additive group of prime order $q$ and generator $P$.
            \item choose $\mathcal{H},$ where $\mathcal{H}$ is a full-domain hash function onto $\mathbb{G}^{2}$, i.e., $\mathcal{H}: \{0,1\}^* \rightarrow \mathbb{G}^2$.
        \end{itemize}
        
        Then, the public parameters $\{ \, \mathbb{G} , q, P, \mathcal{H}\}\,$ are published.


\subsubsection{Smart Meters' Encryption Keys}
KDC generates SMs' encryption keys: $\boldsymbol s_{i} \in$ $\mathbb{Z}_{q}^{2}$, where $\boldsymbol s_{i}$ is the secret key of $SM_i$, for $1 \leq i \leq |\mathbb{SM}|$, and $|\mathbb{SM}|$ denotes the number of SMs in an AMI network.

\begin{table}[!t]
	\centering
	\caption{Main notations.}
	\label{tab:multi_notation}
	
	\begin{tabular}{cl}
		
		\hline
		Notation			& 	Description                                 				\\
		\hline
		\\[-0.7em]
		$SM_i$				&	$i$-th smart meter				            \tabularnewline \\[-0.7em]
 		$\mathit{T_B}$				&	Reporting period used for billing			            \tabularnewline \\[-0.7em]
 		\multirow{2}{*}{$\mathit{T_D}$}				&	Reporting period used for electricity \\& theft detection \tabularnewline \\[-0.7em]
		$\boldsymbol{r}_i[t]$		&	Consumption reading of $SM_i$ at time slot $T_t$				    \tabularnewline \\[-0.7em]
		 \multirow{2}{*}{$\boldsymbol r_{i}$}     & Input (power consumption readings) of the \\ & $SM_i$ over $\mathit{T_D}$  		      \tabularnewline \\[-0.7em]
		$C_i[t]$		&	encrypted reading of $SM_i$ at time slot $T_t$			    \tabularnewline \\[-0.7em]
		\multirow{2}{*}{$\mathbold{c}_{t}$}		& Encrypted consumption report vector from \\& all SMs, $\mathbold{c}_{t} = [C_1[t], C_2[t], \dots, C_{|\mathbb{SM}|}[t] ]$			    \tabularnewline \\[-0.7em]
        $\boldsymbol s_{i}$		&	Secret key of $SM_i$				    \tabularnewline \\[-0.7em]
		 $\ell_t$     & Time slot identifier  		      \tabularnewline \\[-0.7em]
 		$b$				&	Number of readings per billing period			            \tabularnewline \\[-0.7em]
 		\multirow{2}{*}{$d$}			&	Number of readings per electricity \\&  theft detection period			            \tabularnewline \\[-0.7em]
		\multirow{2}{*}{$\mathbold{c}_i^B$}		& Encrypted consumption report vector of $SM_i$ \\                                          & over billing period $\mathit{T_B}$			    \tabularnewline \\[-0.7em]
		\multirow{2}{*}{$\mathbold{c}_i^D$}		& Encrypted consumption report vector of $SM_i$ \\                                          & over electricity theft detection period $\mathit{T_D}$			    \tabularnewline \\[-0.7em]
		$n$				&	Number of neurons in the first hidden layer			            \tabularnewline \\[-0.7em]
 		\multirow{2}{*}{$\mathbold{v}$}				&	Bias vector of size $n$ for the first hidden layer \\ &$\mathbold{v} = [v[1], v[2], \dots, v[n]]$			            \tabularnewline \\[-0.7em]
		$h_i$				&	$i$-th hidden layer			            \tabularnewline \\[-0.7em]
        $\mathbb{G}, q, P$ &   Public parameters for the functional encryption \tabularnewline
          \\[-0.7em]
        $\mathcal{H}: \{0,1\}^* \rightarrow \mathbb{G}^2$        &   Full-domain hash function onto $\mathbb{G}^{2}$
        \tabularnewline \\[-0.7em]
        \multirow{2}{*}{$\boldsymbol{dk_{m}}, DK_{b}, DK_{d}$ }      &    Functional decryption keys for monitoring,  \\     & billing, and electricity theft detection			\tabularnewline\\[-0.7em]
        $\boldsymbol{y_1}, \boldsymbol{y_2}$      &    Vectors corresponding to monitoring and billing    \tabularnewline\\[-0.7em]
        \multirow{2}{*}{$\boldsymbol{W}$}      &    Weights of the first hidden layer \\ &corresponding to electricity theft detection				    \tabularnewline\\[-0.7em]
        $(\boldsymbol{x} \cdot \boldsymbol{y})$        &  Inner/dot product between vectors $\boldsymbol{x}$ and $\boldsymbol{y}$
        \tabularnewline \\[-0.7em]
        
        \hline
	\end{tabular}
\end{table}



\subsubsection{SO's Functional Decryption Keys}
$\boldsymbol{dk_{m}}$, $DK_{b}$, and $DK_{d}$ are the functional decryption keys set used for monitoring, billing, and electricity theft detection, respectively. 
The KDC generates these functional decryption keys as follows.
\renewcommand{\theenumi}{\roman{enumi}}
\begin{enumerate}
    \item Generating $\boldsymbol{dk_{m}}$: A vector of ones, $\boldsymbol{y_1}$, with a length that equals to the number of SMs in an AMI network, is used by the KDC to compute the monitoring functional decryption key $\boldsymbol{dk_{m}}$. This key is sent to the SO such that it can aggregate all the power consumption readings from all SMs at each time slot $T_t$. This vector of ones is used so that when inner product is done with the SMs' readings, the aggregated reading is obtained. The generation of $\boldsymbol{dk_{m}}$ key is as follows.
    
        
        \begin{itemize}
        \item The KDC performs the following operation to compute the monitoring functional decryption key using the SMs' secret keys and $\boldsymbol{y_1}$:
        $$
              \boldsymbol{dk_{m}} = \sum_{i=1}^{|\mathbb{SM}|} \boldsymbol s_{i}  \boldsymbol{y_1}[i]  = \sum_{i=1}^{|\mathbb{SM}|} \boldsymbol s_{i} \in \mathbb{Z}_{q}^{2},
        $$
        where $\boldsymbol{y_1}[i]$ is the $i^{th}$ element in $\boldsymbol{y_1}$. Then the KDC sends the $\boldsymbol{dk_{m}}$ to the SO.
        \end{itemize}
    
    \item Generating $DK_{b}$: The SO sends a vector $\boldsymbol{y_2}$, with a length $b$, to the KDC, where $b$ is the number of readings per billing period $\mathit{T_B}$ as shown in Fig.~\ref{fig:System_scenario}. This vector represents the pricing rates the SO sets and it is used for billing following dynamic pricing approach, i.e., each element in $\boldsymbol{y_2}$ is the electricity rate for one consumption time slot. This allows the SO to compute the inner product operation between $\boldsymbol{y_2}$ and the power consumption of each consumer at different time slots. Using $\boldsymbol{y_2}$, the KDC generates a billing functional decryption key for each $SM_i$ for each billing period as follows.
        
        \begin{itemize}
        \item The KDC calculates the following operations for each $SM_i$ using the $SM_i$'s secret key, $\boldsymbol{y_2}$, and a set of time slot identifiers $\{\ell_1,\ell_2,\dots,\ell_b\}$, as follows:
        $$
        {U}_{\ell t}=\mathcal{H}(\ell_t) \in \mathbb{G}^{2}, 1\leq t \leq b.$$
$$        
        DK_{bi} = \sum_{t=1}^{b} \boldsymbol{y_2}[t] (\boldsymbol{s}_{i}^{\top} \cdot {U}_{\ell t} ) \in \mathbb{G}  
        , 
        $$
        where $( \cdot )$ is the inner/dot product operation between two vectors, and $\boldsymbol{s}_{i}^{\top}$ is the transpose of $\boldsymbol{s}_{i}$.
        \item Next, the KDC sends the $|\mathbb{SM}|$ billing functional decryption keys $DK_{bi}$ to the SO, where $\{1 \leq i \leq |\mathbb{SM}|\}$ and $|\mathbb{SM}|$ is the number of SMs.
        \end{itemize}
    
    \item Generating $DK_{d}$: Regarding the evaluation of electricity theft detection model at the SO-side, the SO sends the first layer's weights of the model ($\boldsymbol{W}$) to the KDC. Supposing that $\boldsymbol{W}$'s dimension is $d$ rows $\times$ $n$ columns, where $d$ is the number of readings per electricity theft detection period $\mathit{T_D}$=$\{\mathit{T_1},\mathit{T_2},\dots,\mathit{T_d}\}$, while $n$ is the number of neurons in the first hidden layer in the model. Then, $\boldsymbol{W}$ can be represented as:
        \begin{center}
       \begin{equation}\label{Y3}
         \boldsymbol{W}=  
        \begin{bmatrix}
            \boldsymbol{w}_1[1] & \boldsymbol{w}_2[1] & \dots & \boldsymbol{w}_n[1]\\
            \boldsymbol{w}_1[2] & \boldsymbol{w}_2[2] & \dots & \boldsymbol{w}_n[2]\\
            . & . & \dots & .\\
            . & . & \dots & .\\
            \boldsymbol{w}_1[d] & \boldsymbol{w}_2[d] & \dots & \boldsymbol{w}_n[d]
        \end{bmatrix},
       \end{equation}
       \end{center}  
       where $\boldsymbol{W}$ is a 2-dimensional array and can be represented as $\boldsymbol{W}=[\boldsymbol{w_1}^{\top},\boldsymbol{w_2}^{\top},\dots,\boldsymbol{w_n}^{\top}]$, $\boldsymbol{w}_i$ is the $i_{th}$ column of $\boldsymbol{W}$, $\boldsymbol{w}_i=[\boldsymbol{w}_i[1],\boldsymbol{w}_i[2],\dots,\boldsymbol{w}_i[d]]^{\top}$ and $\boldsymbol{w}_i \in \mathbb{Z}_{q}^{d}$.
       Therefore, the KDC generates $n$ functional decryption keys corresponding to each column of $\boldsymbol{W}$. In our solution, $\boldsymbol{W}$ is the same for all SMs, i.e., the SO applies a general model to all SMs. Next, the KDC calculates the electricity theft detection functional decryption keys for each SM for each electricity theft detection period as follows.
        
        \begin{itemize}
        \item For each $SM_i$, the KDC performs the following operation using the $SM_i$'s secret key, a set of time slot identifiers $\{\ell_1,\ell_2,\dots,\ell_d\}$, and each column $j$ of $\boldsymbol{W}$, where $j=\{1,  \ldots, n$\}:
        $$
        {U}_{\ell t}=\mathcal{H}(\ell_t) \in \mathbb{G}^{2}, 1\leq t \leq d.$$
        $$
            D_{dji} = \sum_{t=1}^{d} \boldsymbol{w}_j[t] (\boldsymbol{s}_{i}^{\top} \cdot {U}_{\ell t} ) \in \mathbb{G}. 
        $$ 
        \item Next, the KDC sends the $n$ electricity theft detection functional decryption keys to the SO for each $SM_i$:
        $$
        DK_{di}= \{D_{dji}\}_{\forall{j}}.
        $$ 
        \end{itemize}
\end{enumerate}


\subsection{Reporting Fine-grained Power Consumption Readings} \label{sub:Report Generation}
The consumers' fine-grained electricity consumption readings are encrypted by using secret keys sent by the KDC. The SMs transmit the encrypted readings periodically to the SO for load monitoring, billing, and electricity theft detection. 
For each reporting period $T_t$, each $SM_i \in \mathbb{SM}$ generates a power consumption report by executing the following operations.

\begin{itemize}
     \item Each $SM_i$ uses its encryption key $\boldsymbol s_{i}$ and the time slot identifier $\ell_t$ to encrypt its reading $\boldsymbol{r}_i[t]$ in time slot $T_t$ as follows:
      \begin{equation}\label{Enc_eq}
        C_i[t]=(\boldsymbol{s}_{i}^{\top} \cdot {U}_{\ell t} )+\boldsymbol{r}_i[t]P \in \mathbb{G},
  \end{equation}
  where: ${U}_{\ell t}=\mathcal{H}(\ell_t) \in \mathbb{G}^{2}.$
     
\end{itemize}


\subsection{Aggregating Fine-grained Power Consumption Readings for Monitoring}\label{sub:ETDFE_aggregation}

After collecting all the SMs' encrypted readings ($\mathbold{c}_{t}$) at reporting period $T_t$, where $\mathbold{c}_{t} = [C_1[t], C_2[t], \dots, C_{|\mathbb{SM}|}[t] ]$, the SO uses the monitoring functional decryption key $\boldsymbol{dk_{m}}$ to obtain the total aggregated reading for load monitoring by performing the following steps.

 \begin{itemize}
 \item Given the functional decryption key $\boldsymbol{dk_{m}}$ and ciphertexts $\mathbold{c}_{t}$, the SO can compute:
 $$
        {U}_{\ell t}=\mathcal{H}(\ell_t) \in \mathbb{G}^{2}.    
 $$
 
 \item Next, the SO computes:
\begin{equation}\label{decryption_monitoring}
\begin{aligned}{} &\sum_{i=1}^{|\mathbb{SM}|} C_i[t]  -{U}_{\ell t}^{\top}  \boldsymbol{dk_{m}}\\&=\sum_{i=1}^{|\mathbb{SM}|} ((\boldsymbol{s}_{i}^{\top} \cdot {U}_{\ell t} )+\boldsymbol{r}_i[t] P) - (\sum_{i=1}^{|\mathbb{SM}|} \boldsymbol s_{i})^{\top}  {U}_{\ell t}   \\ &=  (\sum_{i=1}^{|\mathbb{SM}|} \boldsymbol s_{i})^{\top}  {U}_{\ell t} +\sum_{i=1}^{|\mathbb{SM}|} \boldsymbol{r}_i[t]   P -(\sum_{i=1}^{|\mathbb{SM}|} \boldsymbol s_{i})^{\top}  {U}_{\ell t}  \\&= (\sum_{i=1}^{|\mathbb{SM}|} \boldsymbol{r}_i[t]) P \in \mathbb{G}.  \end{aligned}
\end{equation}
  \item Finally, the SO uses an approach to compute a discrete logarithm to obtain: $$\sum_{i=1}^{|\mathbb{SM}|} \boldsymbol{r}_i[t] .$$ 
\end{itemize}
In this case, the discrete logarithm is not a difficult problem because $(\sum_{i=1}^{|\mathbb{SM}|} \boldsymbol{r}_i[t] )$ is not a large value. While many methods have been introduced to compute the discrete logarithm such as Shank's baby-step giant-step algorithm~\cite{10.1007/3-540-69053-0_18}, we resorted to using a lookup table to compute it efficiently in a light-weight manner.

By performing the above steps, the result $(\sum_{i=1}^{|\mathbb{SM}|} \boldsymbol{r}_i[t] )$ is 
the summation of the power consumption readings of all SMs at each time slot $T_t$. 
Therefore, ETDFE can achieve the functional requirement (F1) of reporting aggregated power consumption reading for load monitoring by the SO without being able to learn the individual readings to preserve consumers' privacy.

Beside the aforementioned steps, the SO should store the ciphertexts of each $SM_i$ in vector $\mathbold{c}_i^B$ for calculating bills over each billing interval $\mathit{T_B}$ as will be explained in section~\ref{billing}, where $\mathbold{c}_i^B$ is:
\begin{equation*}
		\mathbold{c}_i^B = [C_{i}[1], \dots ,C_{i}[b]]^{\top}.
\end{equation*}

Also, the SO should store the reports of each $SM_i$ over electricity theft detection interval $\mathit{T_D}$
in vector $\mathbold{c}_i^D$ to be applied to the electricity theft detector, at the end of each electricity theft detection interval, as will be explained in section~\ref{sub:PPETD}, where $\mathbold{c}_i^D$ is defined as follows:
\begin{equation*}
		\mathbold{c}_i^D = [C_{i}[1], \dots ,C_{i}[d]]^{\top}.
\end{equation*}


\subsection{Bill Computation Using Dynamic Pricing}\label{billing}
In addition to using the fine-grained power consumption readings in load monitoring and energy management, they are also used to compute bills following dynamic pricing in which the electricity tarrifs are higher in the peak-load periods to stimulate consumers to shift their consumption to off-peak hours to balance electricity supply and demand. In this section, we explain how the SO uses the encrypted power consumption readings to compute bills following dynamic pricing approach.


After collecting $b$ encrypted readings ($\mathbold{c}_i^B$ vector) from each $SM_i$, $\{1 \leq i \leq |\mathbb{SM}|\}$, the SO computes the bill at the end of each billing interval by using the billing functional decryption key $DK_{bi}$ by calculating

\begin{equation}\label{decryption_billing}
\begin{aligned}{} &\sum_{t=1}^{b} \boldsymbol{y_2}[t] C_i[t] - DK_{bi}\\&=\sum_{t=1}^{b} \boldsymbol{y_2}[t] ((\boldsymbol{s}_{i}^{\top} \cdot {U}_{\ell t} )+\boldsymbol{r}_i[t] P) -\sum_{t=1}^{b} \boldsymbol{y_2}[t] (\boldsymbol{s}_{i}^{\top} \cdot {U}_{\ell t} ) \\ &=\sum_{t=1}^{b} \boldsymbol{y_2}[t] (\boldsymbol{s}_{i}^{\top} \cdot {U}_{\ell t} )  +\sum_{t=1}^{b} \boldsymbol{y_2}[t] \boldsymbol{r}_i[t] P- \sum_{t=1}^{b} \boldsymbol{y_2}[t] (\boldsymbol{s}_{i}^{\top} \cdot {U}_{\ell t} ) \\&= (\sum_{t=1}^{b} \boldsymbol{y_2}[t] \boldsymbol{r}_i[t])P. \end{aligned}
\end{equation}

Hence, the SO uses an approach to compute a discrete logarithm to obtain: $$\sum_{t=1}^{b} \boldsymbol{y_2}[t] \boldsymbol{r}_i[t].$$

This is the inner product of the $SM_i$'s power consumption readings and the pricing rates' vector ($\boldsymbol{y_2}$), which is equivalent to the weighted summation of the power consumption readings.
Therefore, ETDFE can achieve the functionality requirement (F2) of computing each consumer's bill following  dynamic prices.


\subsection{Electricity Theft Detection} 
In this section, the dataset used for training the electricity theft detection model is presented, then we explain how we train the model as well as its architecture, and finally, we discuss how the SO can detect electricity thefts without violating the consumers' privacy, i.e., without learning the fine-grained power consumption readings.

\subsubsection{Dataset}
A real smart meter dataset from the Irish Smart Energy Trials \cite{dataset} is used for training and evaluating our electricity theft detector. This dataset was produced in January 2012 by the Electric Ireland and Sustainable Energy Authority of Ireland. It contains electricity consumption readings for more than $1000$ consumers over $536$ days from $2009$ to $2010$, in which an electricity consumption reading is reported by each SM every 30 minutes.  
In our experiment, we used the electricity consumption readings for $|\mathbb{SM}|= 200$ SMs from the dataset. By pre-processing this data, we build $107,200$ records, where each record corresponds to readings of one SM in a single day (i.e., 48 readings).
We define a set $\boldsymbol{r}_i$ of electricity consumption readings (i.e., a record) that are reported by $SM_i$ in each day.
We assume that each electricity theft detection interval is one day, SO the input size of our FFN ($d$) is $48$. 

\textbf{Electricity Theft Attacks:} All the readings in the dataset are for honest consumers.
Although we need to train our model using both honest and malicious data, it is difficult to collect false readings sent by fraudulent consumers. To solve this problem, we created malicious dataset by using a set of electricity theft attacks which are presented in~\cite{jokar2016electricity}.
We considered three types of attacks: by-pass filters, partial reduction, and price-based load control, as summarized in Table \ref{tab:cyberattacks-in}. For each day, $\boldsymbol{r}_i[j]$ denotes the $j^{th}$ electricity reading of $SM_i$. As can be seen in Table \ref{tab:cyberattacks-in}, each function $f(\cdot)$ aims at reducing the power consumption reading $\boldsymbol{r}_i[j]$ by applying different attack scenarios. 
The first attack's objective, i.e., $f_{1}(\cdot)$, is to reduce $\boldsymbol{r}_i[j]$ by a flat reduction ratio $\alpha$, where $0<\alpha<1$, while the attack $f_{2}(\cdot)$ dynamically reduces the reading $\boldsymbol{r}_i[j]$ by a value controlled by the time $\beta[j]$, where $0<\beta[j]<1$. The third attack $f_{3}(\cdot)$ reports the predicted value (mean value) $\mathbb{{E}}[\boldsymbol{r}_i]$ of a fraudulent consumer's power consumption readings for a given day. On the other hand, the fourth attack $f_{4}(\cdot)$ is a By-pass attack, in which the fraudulent consumer sends zero readings during a certain interval (i.e., $[t_{\text{\scriptsize{s}}},t_{\text{\scriptsize{f}}}]$), otherwise, it reports the actual consumption reading $\boldsymbol{r}_i[j]$, where $t_{\text{\scriptsize{s}}}$ and $t_{\text{\scriptsize{f}}}$ are the start and end of the electricity theft interval, respectively.
Similar to the attack $f_{3}(\cdot)$, the attack $f_{5}(\cdot)$ also uses the predicted value (mean value) $\mathbb{{E}}[\boldsymbol{r}_i]$ of a fraudulent consumer's power consumption readings for a given day. But the difference between them is that the readings are reduced dynamically from time to time using $\beta[j]$ in $f_{5}(\cdot)$, where $0<\beta[j]<1$, while the fraudulent consumer who launch $f_{3}(\cdot)$ reports a fixed value during the day. 
Finally, the attack $f_{6}(\cdot)$ is comparatively smart in reducing the electricity bill as it does not change the actual readings during the day but it reports the higher energy consumption readings during low tariff periods.

\begin{table}[t]
    \def\arraystretch{1.5}
    \centering
    \caption{Cyber attacks in Jokar et al. 2016 \cite{jokar2016electricity}.}
	\label{tab:cyberattacks-in}
    \scalebox{0.98}{
    \begin{tabular}{|c|c|}
    \hline
    \rowcolor[gray]{0.8} 
    \textbf{Attack Type} & \textbf{Cyber attacks in Jokar et al. 2016 \cite{jokar2016electricity}} \\ 
    \hline

\multirow{3}{*}{Partial Reduction}&    $f_{1}(\boldsymbol{r}_i[j])=\alpha \boldsymbol{r}_i[j]$      \\ \cline{2-2}
     &    $f_{2}(\boldsymbol{r}_i[j])=\beta[j] \boldsymbol{r}_i[j]$                     \\ \cline{2-2}
     &        $f_{3}(\boldsymbol{r}_i[j])=\mathbb{{E}}[\boldsymbol{r}_i]$                \\ \hline
          By-pass &      \scalebox{0.96}{
     $f_{4}(\boldsymbol{r}_i[j])=\begin{cases}
        0          & \forall t\in[t_{\text{\scriptsize{s}}},t_{\text{\scriptsize{f}}}] \\
        \boldsymbol{r}_i[j] & \forall                                                                 
        t\notin[t_{\text{\scriptsize{s}}},t_{\text{\scriptsize{f}}}]
        \end{cases}$
                                
        }          \\ \hline
     By-pass/Partial Reduction &        $f_{5}(\boldsymbol{r}_i[j])=\beta[j]\mathbb{{E}}[\boldsymbol{r}_i]$                  \\   \hline
        Price-based Load Control & $f_{6}(\boldsymbol{r}_i[j])= \boldsymbol{r}_i[d-j+1] $                  \\ \hline

    \end{tabular}
    }
\end{table}

                                



\begin{figure*}[t]
\centering 
\includegraphics[width=0.9\textwidth]{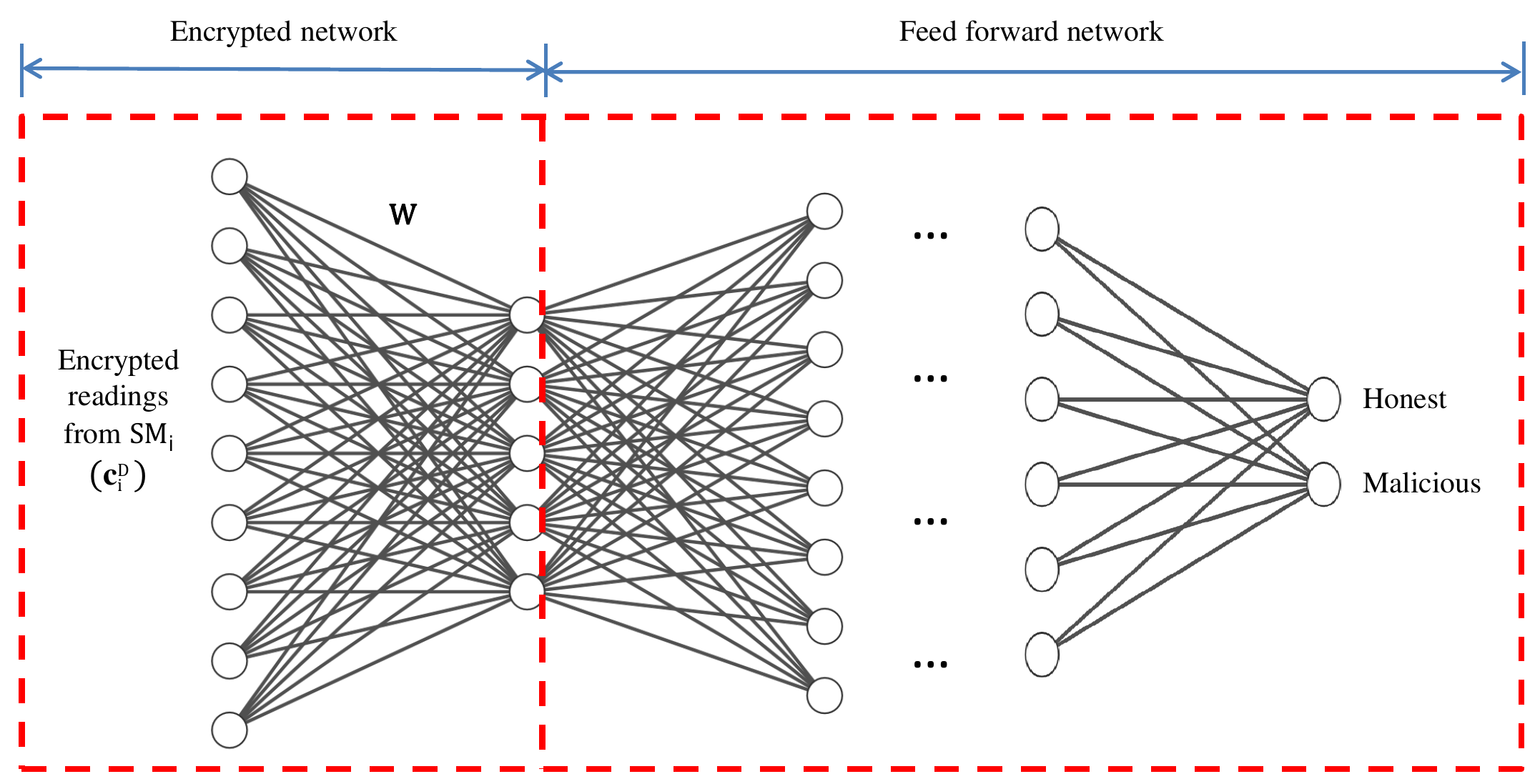}
\caption{Illustration of the encrypted and non-encrypted parts of the proposed theft detection model.} \label{fig:ETD_Model}
\end{figure*}


\textbf{Data Pre-processing:} 
To apply the aforementioned attacks to produce malicious readings, we first set the parameters of each function. 
For functions $f_{1}(\cdot)$, $f_{2}(\cdot)$, and $f_{5}(\cdot)$, $\alpha$ and $\beta[j]$ are random variables that are uniformly distributed over the interval $[0.1,0.6]$ \cite{jokar2016electricity}, while $t_{\text{\scriptsize{s}}}$, in $f_{4}(\cdot)$, is a uniform random variable in $[0,42]$, and the period of the attack, i.e., $t_{\text{\scriptsize{f}}}-t_{\text{\scriptsize{s}}}$, is a uniform random variable in $[6,48]$, and hence, the maximum value of $t_{f}=48$. 
Hence, by applying these attacks on the readings of each SM, the corresponding records for each SM now contains $536$ honest records (for daily readings) and $3,216$ malicious records (i.e., $6 $ $ attacks\times536$). As a result, the dataset is imbalanced because the malicious data is more than the honest data.  

We tackle the problem of imbalanced data by using adaptive synthetic sampling approach (ADASYN) \cite{he2008adasyn} for each SM's records to balance the size of honest and malicious classes.
Thus, each SM has $6,432$ honest and malicious records, where each record contains $48$ electricity consumption readings.
Consequently, the total number of records for $200$ SMs in our dataset is around $1.2$ million. Each SM's records are divided into two datasets for training and testing with the ratio of 4:1. The training datasets are combined together from all SMs to form $\hat{X}_{\text{\sc{tr}}}$ of size about $1$ million records. Similarly, the test datasets are combined together from all SMs to form a test dataset 
of size $257,241$ records. 
Training a model on variety of synthetic attacks' records along with a real dataset~\cite{dataset} helps in improving the model detection rate.


\subsubsection{Electricity Theft Detection Model}
We train a fully connected multi-layer FFN network, i.e., electricity theft detector, with a softmax output layer on $\hat{X}_{\text{\sc{tr}}}$. While training the model, $\ell2-$regularization is used to limit over-fitting, and we adjust the hyper-parmaters of our FFN model using hyperopt tool~\cite{Bergstra_2015} on a validation dataset (which is 33\% of $\hat{X}_{\text{\sc{tr}}}$) to tune the number of neurons in each hidden layer, and select activation function for each layer, batch size, and learning rate. Then, our model is evaluated on the test dataset. In the training phase, Adam optimizer is used to train the model for 60 epochs, 250 batch size, 0.0001 learning rate, and categorical cross entropy as the loss function. To train our model, we used Python3 libraries such as Scikit-learn~\cite{scikit-learn}, Numpy, TensorFlow~\cite{tensorflow2015-whitepaper} and Keras~\cite{chollet2015keras}.
Table~\ref{tab:FFN_arch} gives the detailed  structure of our electricity theft detection model including number of layers, number of neurons, and activation functions.

{\renewcommand{\arraystretch}{1.35}
\begin{table}[t]
\centering
\caption{FFN model used in ETDFE, where AF stands for activation function, and $h_i$ is the $i^{th}$ hidden layer.}
\label{tab:FFN_arch}
\begin{tabular}{| >{\centering\arraybackslash}m{2.2cm}| >{\centering\arraybackslash}m{2.2cm}| >{\centering\arraybackslash}m{2.2cm}|}
\hline \rowcolor[gray]{0.8}
\textbf{Layer} & \cellcolor[gray]{0.8}  \textbf{No. of neurons} & \cellcolor[gray]{0.8}  \textbf{AF} \\ \hline
 Input      &  48    &  Linear\\ \hline
 $h_1$      &  40    &  Linear \\ \hline
 $h_2$      &  500   &  ReLU  \\ \hline
 $h_3$      &  350   &  ReLU  \\ \hline
 $h_4$      &  110   &  ReLU \\ \hline
 $h_5$      &  350   &  ReLU \\ \hline
 $h_6$      &  110   &  ReLU \\ \hline
 $h_7$      &  1536  &  ReLU \\ \hline
 $h_8$      &  500   &  ReLU \\ \hline
 $h_9$      &  1536  &  ReLU \\ \hline
 $h_{10}$   & 500    &  ReLU \\ \hline
 $h_{11}$   & 1536   &  ReLU \\ \hline
 $h_{12}$   & 500    &  ReLU \\ \hline
 $h_{13}$   & 700    &  ReLU \\ \hline
 Output     & 2      &  Softmax\\ \hline
\end{tabular}
\end{table}}

\subsubsection{Privacy-Preserving Evaluation of Theft Detection Model}\label{sub:PPETD}
To enable the SO to evaluate the model we trained without learning the readings to preserve the conumers' privacy, we leverage the inner product operation of FE. As shown in Fig.~\ref{fig:ETD_Model}, only the operations of the first layer of our model architecture are executed using the encrypted data. The result is known to the SO to use in the operations of the next layer.
Generally, the main operation needed by a neural network's feed-forward layers can be expressed by $\mathbold{z} = \boldsymbol{r}\mathbold{W}+\mathbold{v}$ where $\boldsymbol{r}$ is the previous layer input vector, $\mathbold{W}$ is the weight matrix, and $\mathbold{v}$ is the bias vector.
In our FFN model, the weight matrix of the first layer $\boldsymbol{W}$ has dimension ($d \times n$) as shown in Eq.~\ref{Y3}, where $d$ is the number of input neurons (features), and $n$ is the number of neurons in the first hidden layer. In our model, the operation ($\boldsymbol{r}\mathbold{W}+\mathbold{v}$) is performed by multiplying the input vector with $\boldsymbol{W}$, and then the result is added to the bias vector $\mathbold{v}$. This results in $n$ components that are the output of the first hidden layer, which is equivalent to $n$ inner product operations between the input and each column in the weight matrix $\boldsymbol{W}$. Therefore, to preserve the consumers' privacy, we leverage IPFE to do inner product operation on encrypted vectors to obtain the output of the first hidden layer which is:
\begin{equation*}
\boldsymbol r_{i} \boldsymbol{W}+\mathbold{v},
\end{equation*}
where $\boldsymbol{r}_i$ is the input (power consumption readings) of the $SM_i$ over $\mathit{T}_D$, and it can be represented as $[\boldsymbol{r}_i[1], \boldsymbol{r}_i[2],\dots,\boldsymbol{r}_i[d]]$, while $\mathbold{v}$ is the bias vector of size $n$.

After collecting $d$ encrypted readings ($\mathbold{c}_i^D$ vector) from each $\{SM_i, 1 \leq i \leq |\mathbb{SM}|\}$, the SO runs the electricity theft detector by using the functional decryption key $DK_{di}$ corresponding to $SM_i$ to detect whether consumer $i$ is honest or fraudulent. The $n$ columns of $\boldsymbol{W}$ can be represented as $[\boldsymbol{w_1}^{\top},\boldsymbol{w_2}^{\top},\dots,\boldsymbol{w_n}^{\top}]$, where $\boldsymbol{w}_i$ is the $i_{th}$ column of $\boldsymbol{W}$, and $\boldsymbol{w}_i=[\boldsymbol{w}_i[1],\boldsymbol{w}_i[2],\dots,\boldsymbol{w}_i[d]]^{\top} \in \mathbb{Z}_{q}^{d}$.
The evaluation of the electricity theft detection model is done as follows.

 \begin{itemize}
\item Given the functional decryption key $DK_{di}$ and ciphertexts $\mathbold{c}_i^D$ from each $SM_i$ at the end of each electricity theft detection period $\mathit{T_D}$, the SO can compute the inner product between $SM_i$'s ciphertexts $\mathbold{c}_i^D$ and each column of $\boldsymbol{W}$ by performing the following steps. 

\begin{equation}
\begin{aligned}{} &
     \sum_{t=1}^{d} \boldsymbol{w}_j[t] C_i[t]-  D_{dji}\\&
     =\sum_{t=1}^{d} \boldsymbol{w}_j[t] ((\boldsymbol{s}_{i}^{\top} \cdot {U}_{\ell t} )+\boldsymbol{r}_i[t] P) -\sum_{t=1}^{d} \boldsymbol{w}_j[t] (\boldsymbol{s}_{i}^{\top} \cdot {U}_{\ell t} )\\&
    =\sum_{t=1}^{d} \boldsymbol{w}_j[t] (\boldsymbol{s}_{i}^{\top} \cdot {U}_{\ell t} ) +\sum_{t=1}^{d} \boldsymbol{w}_j[t] \boldsymbol{r}_i[t]  P-\\&\sum_{t=1}^{d} \boldsymbol{w}_j[t] (\boldsymbol{s}_{i}^{\top} \cdot {U}_{\ell t} )
= (\sum_{t=1}^{d} \boldsymbol{w}_j[t] \boldsymbol{r}_i[t]) P
\end{aligned}
\end{equation}

\item These equations are computed for $j={1,2,\dots,n}$. The SO uses an approach to compute a discrete logarithm to obtain:
$$\sum_{t=1}^{d} \boldsymbol{w}_j[t] \boldsymbol{r}_i[t]$$



\item The results are in clear form. Then the SO adds them to the bias $\mathbold{v}$ of the first hidden layer to obtain the output of the first hidden layer of the electricity theft detector as follows:
\begin{align*}
 [(\boldsymbol{r_{i}}\cdot \boldsymbol{w_1}^{\top})+\mathbold{v}[1], (\boldsymbol{r_{i}}\cdot \boldsymbol{w_2}^{\top})+\mathbold{v}[2], \dots, \\ (\boldsymbol{r_{i}}\cdot \boldsymbol{w_n}^{\top})+\mathbold{v}[n]]  
 \end{align*}
Then, the output of the first hidden layer is the input to the next layer of the model and the operations of next layers are completed until the calculations are done in the last layer and the classification result is obtained. 
\end{itemize}
 
Note that, the number of neurons in the first hidden layer should be fewer than the number of inputs (i.e., $n < d$) because if $n \geq d$, the SO may obtain the fine-grained readings, since $d$ unknowns in $d$ equations may be solved to obtain the readings.

Therefore, FFN model is evaluated securely by the SO at the end of each electricity theft detection interval without learning the consumption readings to preserve the consumers' privacy. Therefore, ETDFE can achieve the functionality requirement (F3) of privacy-preserving electricity theft detection. 

%% file: Files/Security.tex
\section{Performance Evaluation} \label{sec:Performance Evaluations} 
	
In this section, we first evaluate the performance of the electricity theft detection model, and then assess our scheme in terms of communication and computation overhead.

\subsection{Electricity Theft Detection}

\textbf{Performance Metrics:} In order to evaluate our scheme's performance, we considered the following metrics. The detection rate ($DR$) measures the percentage of fraudulent consumers that are detected correctly. 
The false acceptance rate ($FA$) measures the percentage of the honest consumers that are falsely recognized as fraudulent. The highest difference ($HD$) is the difference between $DR$ and $FA$. The accuracy measures the percentage of honest/fraudulent consumers that are correctly detected as honest/fraudulent. The model performance is better when $DR$, $HD$, and accuracy are high, and $FA$ is low.

\begin{equation*}
    \small
   \textrm{$DR$} =  \frac{\textrm{$TP$}}{\textrm{$TP$} + \textrm{$FP$}}, \quad \textrm{$FA$} = \frac{\textrm{$FP$}}{\textrm{$TN$} + \textrm{$FP$}},
\end{equation*}
\begin{equation*}
    \small
\quad  \textrm{$HD$} =  \textrm{$DR$} - \textrm{$FA$},
 \quad  Accuracy = \frac{\textrm{$TP$}+ \textrm{$TN$}}{\textrm{$TN$} + \textrm{$TP$} + \textrm{$FP$} + \textrm{$FN$}}, 
   \end{equation*}
where, $TP$, $TN$, $FN$, and $FP$ stand for true positive, true negative, false negative, and false positive, respectively.


\textbf{Results and Discussion:} We have evaluated our model using the confusion matrix which is imported from Scikit-learn python library~\cite{scikit-learn}.
Our baseline is the plaintext FFN model (without privacy preservation) and we also compare it with our privacy-preserving model. We compare our results with the proposed scheme in~\cite{jokar2016electricity} and the three models proposed in PPETD~\cite{8746794}, which are MD1 with ``28 CNN filters, 1 stride size, 6 units filter size, and 2,048 hidden units"; MD2 with ``256 CNN filters, 1 stride size,  5 units filter size, 1,536 hidden units"; and MD3 with ``64 CNN filters, 1 stride size, 5 units filter size, 1,536 hidden units". 

{\renewcommand{\arraystretch}{1.35}
\begin{table*}[t]
    \centering
    \caption{The results of our FFN model and other models in literature.}
    \label{tab:results}

    \begin{tabular}{| >{\centering\arraybackslash}m{3cm} | >{\centering\arraybackslash}m{2cm} | >{\centering\arraybackslash}m{2cm} | >{\centering\arraybackslash}m{2cm}| >{\centering\arraybackslash}m{2cm} | >{\centering\arraybackslash}m{2cm} |
}
    \hline
    \rowcolor[gray]{0.8} 
    \textbf{Model} & \textbf{Method} & \textbf{DR(\%)} & \textbf{FA(\%)} & \textbf{HD(\%)} & \textbf{Accuracy(\%)} \\ \hline
    & Without privacy preservation   &   92.56   &   5.84 &   \textbf{86.72}  &  \textbf{93.36}  \\ \cline{2-6} 
    \multirow{-2}{*}{\textbf{ETDFE}} & With privacy preservation   &  \textbf{92.56}  &    5.84  & \textbf{86.72}        & \textbf{93.36}  \\ \hline
    & Without privacy preservation &   93.6   &   8.00  &   85.6 &  93.2 \\ \cline{2-6} 
    \multirow{-2}{*}{\textbf{PPETD MD1~\cite{8746794}}}  & With privacy preservation&  91.5    &  7.40  &   84.1  & 91.8 
    \\ \hline
    & Without privacy preservation &    92.9 &  8.80 &  84.0  &  92.4  \\ \cline{2-6} 
    \multirow{-2}{*}{\textbf{PPETD MD2~\cite{8746794}}} & With privacy preservation & 90.0  &  8.79 &  81.2  & 90.2 \\ \hline
    & without privacy preservation &  91.5  &4.80 &  86.7  &  92.4  \\ \cline{2-6} 
    \multirow{-2}{*}{\textbf{PPETD MD3~\cite{8746794}}} & With privacy preservation&  88.6 & 3.90 & 84.6 &   90.3\\ \hline
    \textbf{Jokar et al. 2016 \cite{jokar2016electricity}} & without privacy preservation &  94.0 &  11.0 &  83.0 & --    \\ \hline
    \end{tabular}
\end{table*}}


Table~\ref{tab:results} provides the evaluation results for our proposed model and the existing models in the literature with and without privacy preservation. Considering privacy-preserving electricity theft detection, our scheme ETDFE offers higher accuracy and $DR$, 93.36\% and 92.56\%, respectively, compared to PPETD MD1~\cite{8746794} which has 91.8\% accuracy and 91.5\% $DR$. PPETD MD3~\cite{8746794} has the lowest $FA$ which equals to 3.9\%, while ETDFE has 5.84\%, which is slightly higher but it is still acceptable. Furthermore, unlike~\cite{jokar2016electricity} that creates one model for each consumer, our detector is a general model that does not rely on specific consumer's data, and can be applied to new consumers who have no history of power consumption. 
Moreover, our scheme has higher $HD$ than~\cite{jokar2016electricity} and~\cite{8746794}.

\begin{figure}[t]
\centering
\includegraphics[width=0.45\textwidth]{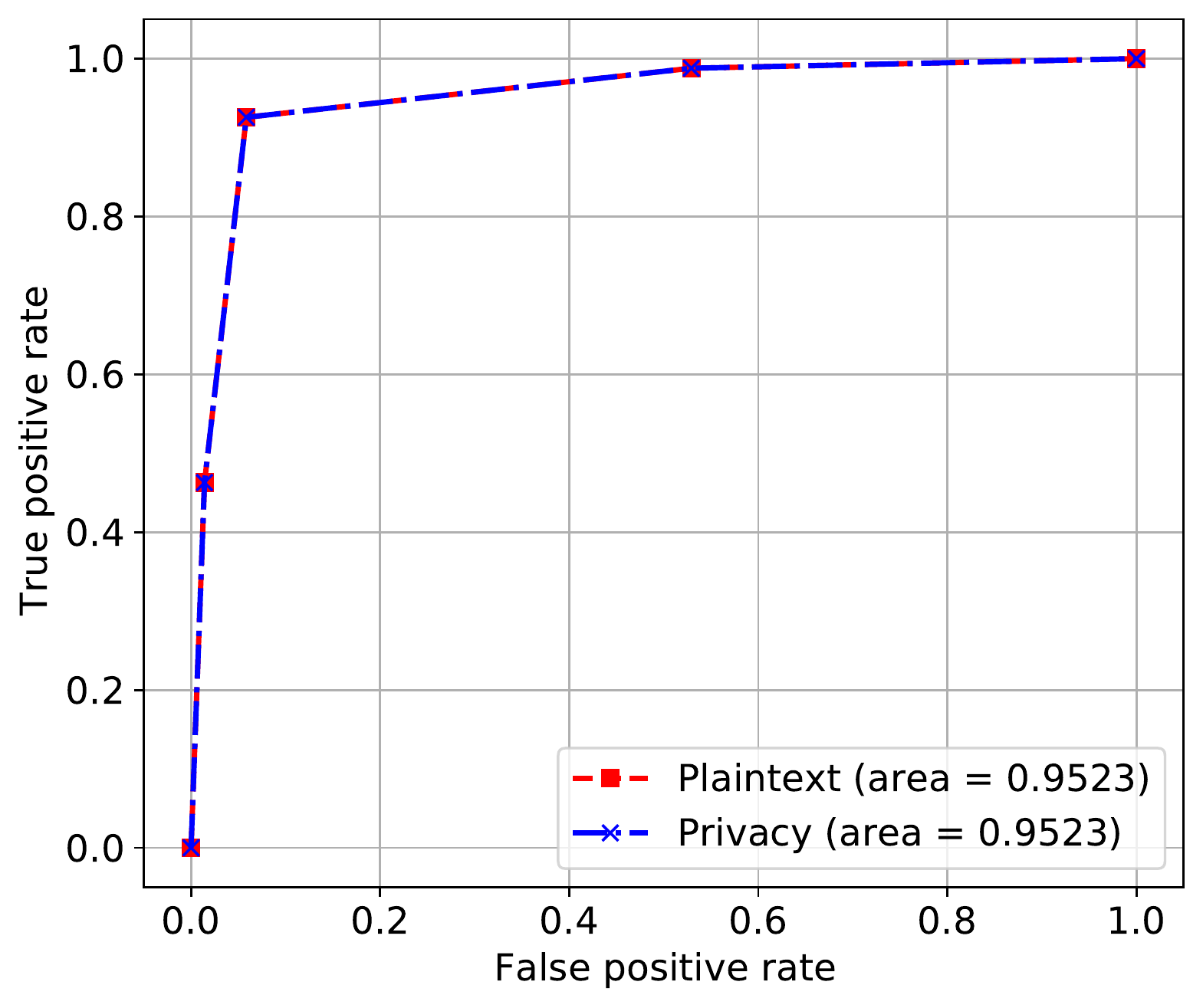}
\caption{ROC of our model, for the plaintext model (without privacy preservation) and our privacy-preserving model.} \label{fig:roc}
\end{figure}

In addition, Fig.~\ref{fig:roc} shows the Receiver Operating Characteristics (ROC) curves for our model with and without privacy preservation. ROC curve is often used to evaluate the classification accuracy, which is measured by the area under the ROC curve (AUC). This area indicates how much the model can distinguish between the classes, where a higher AUC represents a better performance.
The given results indicate that the overall accuracy of our model does not degrade when using our privacy-preserving evaluation technique because the results of the inner product of the encrypted vectors using FE is similar to the result of the inner product of the plaintext vectors.
This is different from the proposed scheme in~\cite{8746794} that suffers from accuracy reduction when considering privacy preservation. This reduction occurs because a non-linear function (sigmoid) is approximated as a linear function in order to be able to evaluate the model on masked readings. 


\subsection{Computation and Communication Overhead} 
Our scheme is implemented using Python ``Charm'' cryptographic library~\cite{Akinyele2013} running on a standard desktop computer with an Intel Core $i7$ Central Processing Unit (CPU) operating at 2GHz and 8GB of Random Access Memory (RAM). We used elliptic curve of size 160 bits (MNT159 curve).


\subsubsection{\textbf{Computation Overhead}}
To evaluate ETDFE, we compare our scheme's computation overhead with the one presented in~\cite{8746794} for load monitoring, billing, and electricity theft detection.
For power consumption reporting, the SMs' computation overhead needed to encrypt the power consumption reading by using Eq.~\ref{Enc_eq} is 0.009 ms, compared to 0.35 ms in PPETD~\cite{8746794}, as can be seen in Table~\ref{computation_overhead}. The results confirm that the computation overhead on the SMs is low which is important because the SMs are resource-constrained devices.
On the other hand, the overhead of aggregating $200$ readings by the SO is 47.2 $\mu$s using our scheme, while it is 0.071 $\mu$s in PPETD. Although our scheme need more time for aggregating the SMs' readings, it is still low.
Therefore, the comparison with PPETD demonstrates that our ETDFE scheme can reduce the computation overhead of reporting a power consumption reading on SMs by 97.4\%. 

For our privacy-preserving FFN model evaluation, the total time needed to evaluate 8,318 hidden units over 15 layers FFN model is around 0.82 seconds for each consumer at the end of the electricity theft interval, while PPETD requires 48 minutes to evaluate the model. Therefore, our scheme provides 99.9\% improvement in evaluating the electricity theft detection model by the SO. It is worth nothing that this 0.82 seconds includes the decryption of the first layer and obtaining the result of the classifier. 
Moreover, unlike PPETD, our scheme does not need each SM and the SO to be engaged in online/interactive session for evaluating the electricity theft detection model.

\subsubsection{\textbf{Communication Overhead}}
We used elliptic curve, in the cryptography operations needed for our scheme, which provides 160 bits security level. As can be seen from Eq.~\ref{Enc_eq}, each SM sends an encrypted fine-grained reading of total size of 40 bytes. For privacy-preserving evaluation of electricity theft detection model, the SO uses the stored ciphertexts sent by each SM; therefore, no additional communication overhead is needed between the SO and the SMs. On the other hand, PPETD uses masked readings to preserve consumers' privacy, and also uses secure multiplication, secure evaluation of $sigmoid(.)$, and garbled circuit for privacy-preserving evaluation of a CNN model. This leads to a high communication overhead of around 1900 MB per SM. As a result, Our scheme offers a significantly lower communication overhead in comparison with PPETD. 
\begin{center}
{\renewcommand{\arraystretch}{1.5}
\begin{table}[t]
\centering
\caption{Computation overhead of our scheme and PPETD~\cite{8746794}. \label{computation_overhead}}
\begin{tabular}{|c|c|c|c|}
\hline \cellcolor[gray]{0.8} 
\textbf{Methodology} & \cellcolor[gray]{0.8}  \textbf{Encryption} & \cellcolor[gray]{0.8}  \textbf{Aggregation} &\cellcolor[gray]{0.8} \textbf{Model evaluation}\\ \hline
 ETDFE  &  0.009 ms    &  47.2 $\mu$s & 0.82 s \\ \hline
 PPETD~\cite{8746794}    &  0.35 ms    &  0.071 $\mu$s &  2880 s\\ \hline
\end{tabular}
\end{table}}
\end{center}
\section{Security and Privacy Analysis}\label{sec:Security Analysis}
Our scheme can achieve the following desirable security/privacy requirements that can counter the attacks mentioned in section~\ref{subsec:Threat Model}. 





\textbf{Theft detection:} To ensure the secure evaluation of our electricity theft detector, each SM first encrypts its fine-grained power consumption readings using FE, and then, the SO uses the functional decryption keys to get the output of the first layer without being able to learn the individual readings of the SM. Then, this output can be used to obtain the classification of the model. 
In addition, our scheme ensures that only the SO knows the result of the electricity theft detector, unlike PPETD~\cite{8746794} in which the result is revealed to both the SO and SM. This may give the consumer enough time to change the malicious software of the SM before the SO sends technicians to inspect it to avoid liability.

On the other hand, the SO uses the same encrypted readings for monitoring, billing, and evaluation of the electricity theft detector. Thus, our scheme ensures that a consumer will not be able to fool the detector by sending two readings; one false reading for billing/monitoring and another true reading for theft detection. 
Therefore, our scheme is secured against this misbehaviour, and hence, it can satisfy the security requirement of privacy-preserving theft detection (S1).


    

\textbf{Consumers' privacy preservation:} The consumers' fine-grained power consumption readings are encrypted and no entity (including SO) is able to learn the individual readings to preserve consumers' privacy. In addition, if the same reading is repeated at different times, the ciphertext looks different because each time the encryption is done using different time slot identifier and thus ${U}_{\ell t}$ cannot be repeated. If ${U}_{\ell t}$ is reused, the ciphertexts of two readings of the $SM_i$ ($\boldsymbol{r}_i[1]$ and $\boldsymbol{r}_i[2]$) are: $c_i[1]=(\boldsymbol{s}_{i}^{\top} \cdot {U}_{\ell})+\boldsymbol{r}_i[1]P$ and $c_i[2]=(\boldsymbol{s}_{i}^{\top} \cdot {U}_{\ell})+\boldsymbol{r}_i[2]P$, respectively. Hence, by subtracting the two ciphertexts: $c_i[1] - c_i[2]=(\boldsymbol{r}_i[1]-\boldsymbol{r}_i[2])P$, by knowing one reading, the other can be obtained. To learn a certain consumer's power consumption reading, the SO must collude with ($\mathbb{|SM|}$-1) consumers. This can be done by subtracting the total power consumption of the colluding SMs from the total power consumption known to the SO. This attack is not feasible when the number of SMs in an AMI network is large.
In addition, although the SO has $\sum_{t=1}^{b} \boldsymbol{y_2}[t] (\boldsymbol{s}_{i}^{\top} \cdot {U}_{\ell t})$, ${U}_{\ell t}$, and $\boldsymbol{y_2}$ for the billing process, it is difficult to obtain the $SM_i$'s secret key $\boldsymbol{s_{i}}$ and using it to compute the $SM_i$'s future readings, because ${U}_{\ell t}$ changes, and thus it is infeasible to solve the discrete logarithmic problem.
Therefore, ETDFE satisfies the security requirement of privacy preservation (S2).
	
\textbf{Confidentiality of AMI's total power consumption and  consumers' bills:} After receiving the encrypted fine-grained power consumption readings from SMs, the SO can aggregate the readings to obtain the total power consumption for load monitoring. Attackers, who may be able to intercept the encrypted readings, learn nothing about the total consumption of an AMI because a private key known only to the SO is needed to calculate the aggregated power consumption readings. Also, the SO is the only entity which is capable of computing the bill of each consumer since a secret key known only to the SO is needed. Thus, ETDFE satisfies the security requirement of the aggregated power confidentiality (S3).


%% file: Files/Conclusion.tex
\section{Conclusion} \label{sec:Conclusions}

In this paper, we have proposed ETDFE, a novel scheme that uses encrypted fine-grained power consumption readings reported by the SMs for electricity theft detection, load monitoring, and computation of electricity bills following dynamic pricing while preserving consumers' privacy. 
To preserve privacy, no entity is able to learn the fine-grained power consumption readings of individual consumers. Functional encryption is used by each consumer to encrypt the power consumption readings and the SO uses a functional decryption key to compute bills and total power consumption for load management, and evaluate a machine learning model using a set of encrypted power consumption readings to detect electricity theft. 
Moreover, extensive simulations have been conducted using real dataset to evaluate our scheme. The given results indicate that our scheme can detect fraudulent consumers accurately and preserve consumers' privacy with acceptable communication and computation overhead. 
Unlike~\cite{8746794}, our scheme does not suffer from accuracy degradation due to the privacy-preserving evaluation of the model. 
Furthermore, the comparison with~\cite{8746794} demonstrates that our scheme can reduce the computation overhead of reporting a power consumption reading on SMs by 97.4\%, while offering a significantly lower communication overhead. Unlike~\cite{8746794}, the SO and SMs do not need to establish an online/interactive session to evaluate the electricity theft detection model, and we also reduce the computation and communication overhead from 48 minutes to only 0.82 seconds, and from 1900 MB per SM to only 40 bytes, respectively.

\section*{Acknowledgement}
This project was funded by the Deanship of Scientific Research (DSR) at King Abdulaziz University, Jeddah, under grant no. DF-745-611-1441. The authors, therefore, acknowledge with thanks DSR for technical and financial support.